\documentclass[twocolumns]{IEEEtran}
\usepackage{graphicx}
\usepackage{amsmath}
\usepackage{authblk}
\usepackage{blindtext}
\usepackage{enumitem}
\usepackage[colorlinks=true, linkcolor=blue, urlcolor=blue, citecolor=blue, pdfborder={0 0 0}]{hyperref}
\usepackage[normalem]{ulem} 
\usepackage[utf8]{inputenc}

\usepackage{fancyhdr} 

\fancypagestyle{firstpage}{
  \fancyhf{} 
  \fancyhead[R]{\fontsize{10}{12}\selectfont\thepage} 
}

\title{Secure Wearable Apps for Remote Healthcare Through Modern Cryptography}
\author[1]{Andric Li}
\author[2]{Grace Luo}
\author[3]{Christopher Tao}
\author[4]{Diego Zuluaga}
\affil[1]{Department of Computer Science, UCSD}
\affil[2]{Department of Computer Science, U-Wisconsin-Madison}
\affil[3]{The Bishop’s School}
\affil[4]{Futurewei}

\begin{document}
\maketitle
\thispagestyle{firstpage} 
\begin{abstract}

Wearable devices like smartwatches, wristbands, and fitness trackers are designed to be lightweight devices to be worn on the human body. With the increased connectivity of wearable devices, they will become integral to remote healthcare solutions. For example, a smartwatch can measure and upload a patient's vital signs to the cloud through a network which is monitored by software backed with Artificial Intelligence. When an anomaly of a patient is detected, it will be alerted to healthcare professionals for proper intervention. Remote healthcare offers substantial benefits for both patients and healthcare providers as patients may avoid expensive in-patient care by choosing the comfort of staying at home while being monitored after a surgery and healthcare providers can resolve challenges between limited resources and a growing population.

While remote healthcare through wearable devices is ubiquitous and affordable, it raises concerns about patient privacy. Patients may wonder: Is my data stored in the cloud safe? Can anyone access and manipulate my data for blackmailing? Hence, securing patient private information end-to-end becomes crucial. This paper explores solutions for applying modern cryptography to secure wearable apps and ensure patient data is protected with confidentiality, integrity, and authenticity from wearable edge to cloud.

\end{abstract}
Index Terms- wearable apps, remote healthcare, cryptography, security

\section{Introduction}

While remote healthcare has become a compelling solution with the advent of wearable devices, a patient's privacy must be properly addressed to ensure patient data is protected from any adversaries, from the point of data collection through a wearable or IoT device at the network edge to the point where the data is stored and retrieved by healthcare providers in a public cloud platform. Cloud service providers (CSPs), like Microsoft Azure or Google, provide tools and capabilities to healthcare providers to secure user data after it has been uploaded. However, user privacy at the cloud edge is outside the protection scope of CSPs. The protection of data collected from wearables falls under the responsibility of applications on these devices. Fortunately, modern cryptography adopted by cloud service providers can be leveraged to protect user data collection by smart wearable apps. The US NIST has established cryptographic standards, such as the Advanced Encryption Standard (AES) for data encryption and Diffie-Hellman Key Exchange (DH) for pair-wise key establishment. These cryptographic algorithms facilitate smart wearable apps in establishing a unique secure channel between the edge device and the cloud healthcare database. The secure channel mutually authenticates the patient and the target healthcare database in the cloud. In addition, the secure channel derives a session key for subsequent data transfers. Thus, patient data will be encrypted by a dedicated and unique key at the point of collection and decrypted only at the target cloud database.

This research prototypes a secure smartwatch app for health monitoring in remote patient care. This app illustrates how a secure channel is constructed using crypto APIs provided by Android Studio. It demonstrates how patient data is encrypted at the time of collection from the sensor, before being sent to the remote server through an open network, and how the remote server decrypts the patient data, ensuring confidentiality, authenticity, and integrity.

This paper consists of five sections as following:

\begin{enumerate}
\item Introduction
\item Standards And Government Regulations 
\item Cryptography for Remote Healthcare App
\item Development Platform
\item Implementation
\item Conclusion
\end{enumerate}

\section{Standards And Government Regulations }
The US Health Insurance Portability and Accountability Act (HIPAA) emphasizes technical safeguards, as the standards developed for these represent sound business practices for technology and related procedures and policies.
\cite{hipaaTechSafeguards} As specified in the HIPAA § 164.312(e)(2)(ii):

Where this implementation specification is a reasonable and appropriate safeguard for a covered entity, the covered entity must: “Implement a mechanism to encrypt electronic protected health information whenever deemed appropriate.”

This instruction from HIPAA seems vague and leaves room for interpretation, as the HIPAA rule allows companies to use any security measures that enable them to apply the standards reasonably and appropriately to safeguard electronic protected health information (EPHI). In contrast, the European Union’s General Data Protection Regulation (GDPR) is more specific. In \href{https://eur-lex.europa.eu/legal-content/EN/TXT/HTML/?uri=CELEX:32016R0679&from=EN#d1e3383-1-1}{\uline{Article 32}}, the GDPR specifies:
“...the controller and the processor shall implement appropriate technical and organisational measures to ensure a level of security appropriate to the risk, including inter alia as appropriate 
\cite{gdpr32}:

\begin{enumerate}[label=\alph*)]
    \item pseudonyms and encryption of personal data
    \item the ability to ensure the ongoing confidentiality, integrity, availability and resilience of processing systems and services
    \item the ability to restore the availability and access to personal data in a timely manner in the event of a physical or technical incident
    \item a process for regularly testing, assessing and evaluating the effectiveness of technical and organizational measures for ensuring the security of the processing.
\end{enumerate} 

And in \href{https://www.privacy-regulation.eu/en/recital-83-GDPR.htm}{\uline{Recital 83}}, it further specifies:
“In order to maintain security and to prevent processing in infringement of this Regulation, the controller or processor should evaluate the risks inherent in the processing and implement measures to mitigate those risks, such as encryption. Those measures should ensure an appropriate level of security, including confidentiality, taking into account the state of the art and the costs of implementation in relation to the risks and the nature of the personal data to be protected. ...” \cite{gdprRecital83}

Although the GDPR is aimed at protecting data privacy in the EU nations, the regulation also applies to entities that want to operate within the EU, and thus has broader implications beyond the EU.

\section{Cryptography for Securing remote healthcare applications}

AES-GCM is an authenticated encryption algorithm developed to protect data in transit. AES-GCM encryption produces ciphertext using counter mode AES, along with a tag produced by a Galois field multiplier. Both the ciphertext and the tag are transmitted through an open network. Upon arrival, the receiver decrypts the datagram ciphertext, generates a new tag, and compares it with the received tag. If there is any data tampering, a tag mismatch will detect the anomaly. Subsequently, the driver erases the decrypted content and returns an error status. In a normal case, where the tags match, the driver will render the plaintext.

\begin{center}
    \raisebox{-0.5\height}{\includegraphics[width=7.5cm]{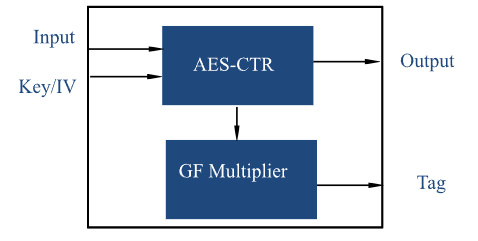}}
\end{center}

AES-GCM requires a session key for data encryption. Session keys can be established using the ECDH protocol, which is a key exchange protocol developed to derive a common session key based on a public/private key pair between two remote entities. Here’s an example illustrating how ECDH works. Assume Alice and Bob want to establish a session key. They agree on a set of predefined domain parameters, including a generator called G.

Alice generates a private key dA using a random number generator, and its corresponding public key QA.

\begin{center} $QA=dA*G$ \end{center}

Bob generates a private key dB using his random number generator, and its corresponding public key QB.

\begin{center} $QB=dB*G$ \end{center}

Alice and Bob exchange public keys QA and QB through an open network. Alice then performs the following operation:

\begin{center} $CA= dA*QB=dA*dB*G$ \end{center}

Bob performs the corresponding operation:

\begin{center} $CB=dB*QA=dB*dA*G$ \end{center}

Note that CA and CB are identical; hence, CA and CB become the session key for Alice and Bob to encrypt and decrypt subsequent messages.







\section{Development platform}




The development platform for this project and its main components are illustrated in the diagram below. Android Studio is used as the development platform for its ease of development and rich platform ecosystem, it is an integrated development environment for developing Android applications \cite{androidStudio}. Android Studio includes an Android Emulator that simulates Android devices on the computer, mimics the hardware and software environment of an Android device, and allows developers to run and test Android applications without needing a physical Android device.

The project instantiated a virtual smartwatch on the Android Emulation platform and created a Smart Watch Wear OS app to run on the virtual device \cite{androidEmulator}. The application implements a heart rate monitor function defined by the Wear OS health service and calls the APIs provided by Wear OS to use platform functions, specifically the secure HTTP communication function in this case \cite{wearOsPairing}.

The Wear OS platform handles network transitions, including the encryption of heart rate information before transmission. The cryptography techniques and the key exchange protocol discussed in the earlier section are used in this process. As shown in the diagram, the key is created by the key manager and stored in the Cloud. Only the metadata, or handle of the key, is shared with the application. Using this handle, Wear OS encrypts the data and shares the handle (a public key) with the receiving side, such as a health center.

For design simplicity, this project was designed without a phone: the smartwatch communicates directly with the server, which is implemented using a Python script running on a local computer to mimic a health center securely receiving heart rate information from the wearable device. Since the heart rate information is transmitted in encrypted format, the server uses the key handle passed from Wear OS to retrieve the key and decrypt the data, thereby obtaining the user's heart rate.


\begin{center}
    \raisebox{-0.5\height}{\includegraphics[width=7.5cm]{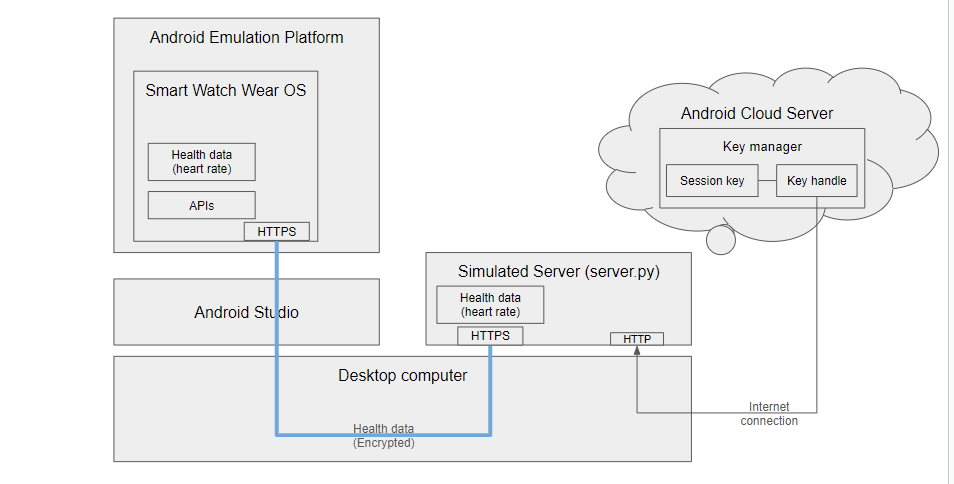}}
\end{center}

\section{Implementation}


First, we use OpenSSL to create an RSA private key and its self-signed public key certificate to form a certificate chain. The certificate chain is bound to the SSL context as follows. The "server.pem" is the private key, and "certificate.crt" is the public key.
\begin{center}
    \raisebox{-0.5\height}{\includegraphics[width=7.5cm]{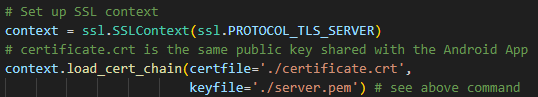}}
\end{center}

The same public certificate is then distributed to the client device. The client’s "network\_security\_config.xml" identifies the public key certificate directory as shown below. This public key certificate must be placed in the listed certificate directories.
\begin{center}
    \raisebox{-0.5\height}{\includegraphics[width=7.5cm]{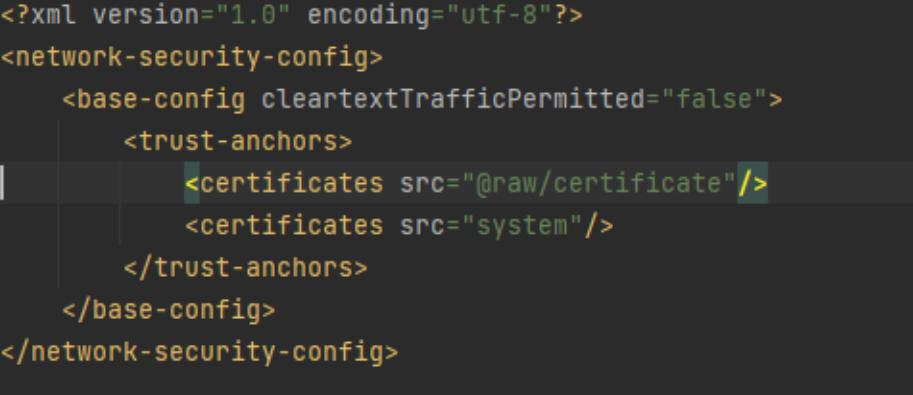}}
\end{center}

Next, we define the naming scheme in the httpUrl variable in the client app as "https://". This allows the HTTP protocol to implicitly recognize the security scheme.

\begin{center}
    \raisebox{-0.5\height}{\includegraphics[width=7.5cm]{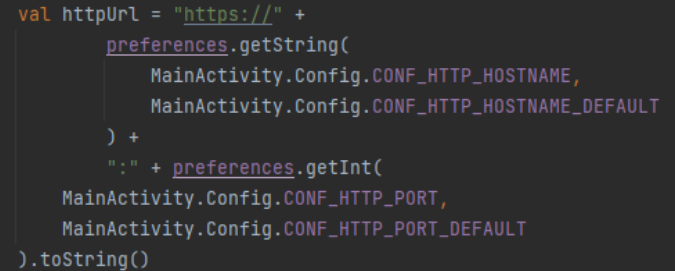}}
\end{center}

This setup effectively establishes a secure communication channel between the server and client using public key cryptography. Subsequent operations, such as session key establishment, payload encryption, and integrity protection, are encapsulated and handled by the SSL protocol.


\section{Conclusion}

The simulation demonstrated data protection on the secure smartwatch for health monitoring in remote patient care. This application illustrates how a secure channel is constructed using crypto APIs provided by Android Studio, how patient heart rate data is encrypted using the Wear OS API before being sent to the remote server through an open network, and how the remote server decrypts the patient's heart rate data with assurance of confidentiality, authenticity, and integrity.

This simulation established the secure channel at the time of the smartwatch-to-server connection and performed encryption and decryption for every data transfer. Our demo shows that HTTPS is a viable solution for Wear OS applications and fulfills the goals as described in the HIPAA security standard § 164.312(e)(2)(ii): “Adopting a single industry-wide encryption standard in the Security Rule would likely have placed too high a financial and technical burden on many covered entities. The Security Rule allows covered entities the flexibility to determine when, with whom, and what method of encryption to use.”

\section{Acknowledgements}

We would like to express our gratitude to Preston Lau for his invaluable suggestions during the incubation and planning of this project. His willingness to give his time and share his insights is greatly appreciated.

\bibliographystyle{ieeetr}
\bibliography{Bibliography}
\end{document}